\documentclass[%
reprint,
nofootinbib,
 amsmath,amssymb,
 aps,
pra,
floatfix,
]{revtex4-1}

\usepackage{float}
\makeatletter
\let\newfloat\newfloat@ltx
\makeatother

\usepackage{algorithmicx}
\usepackage{algorithm}
\usepackage[noend]{algpseudocode}
\usepackage{bm}
\usepackage{color}
\usepackage{amssymb}
\usepackage{amsmath}
\usepackage{graphicx}
\usepackage[outdir=./figs/]{epstopdf}
\usepackage{comment}
\usepackage[parfill]{parskip}
\usepackage{enumitem}
\usepackage{verbatim}
\usepackage{tikz}
\usepackage{mathtools}
\usepackage{colortbl}

\newcommand{\hp}{H_{\mathrm{p}}}
\newcommand{\Ds}{D_{\mathrm{q}}}
\newcommand{\Dpw}{D_{\mathrm{pw}}}
\newcommand{\hd}{H_{\mathrm{d}}}
\newcommand{\hz}{H_z}

\newcommand{\kett}[1]{| #1\rangle}
\newcommand{\braa}[1]{\langle #1 |}

\begin{document}

\title{Achieving fair sampling in quantum annealing}

\author{Vaibhaw Kumar$^1$, Casey Tomlin$^1$,
Curt Nehrkorn$^1$,
Daniel O'Malley$^2$, Joseph Dulny III$^1$}

\affiliation{%
  $^1$Booz Allen Hamilton\\
    $^2$Los Alamos National Laboratory\\
}

\date{\today}

\begin{abstract}
Sampling all ground states of a Hamiltonian with equal probability is a desired feature of a sampling algorithm, but recent studies indicate that common variants of transverse field quantum annealing sample the ground state subspace unfairly. In this note, we present perturbation theory arguments suggesting that this deficiency can be corrected by employing reverse annealing-inspired paths. We confirm that this conclusion holds in simulations of previously studied small instances with degeneracy, as well as larger instances on quantum annealing hardware.
\end{abstract}

\maketitle

Quantum annealing (QA)-based optimization has gained a lot of interest recently, fueled partially by the advent of QA hardware \cite{bian2010ising}. The usual aim of this technique is to produce a ground state of a Hamiltonian cost function with sufficiently high probability, with less attention paid to the distribution of sampled solutions. In cases where there are several cost function minima, one would hope to find that each is sampled with equal probability across many annealing runs; i.e., one would like a protocol that achieves \emph{fair sampling}. Fair sampling is crucial for a number of applications. One such example, the \emph{set membership problem}, ubiquitous in computer science \cite{broder2004network, radvilavicius2012overview,tarkoma2011theory}, involves finding whether an element belongs to a given set. SAT-based probabilistic membership filters \cite{weaver2012satisfiability, 10.1145/800133.804350, douglass2015constructing, fang2018nae, azinovic2017assessment} used to establish set membership require access to not one, but a majority of the solutions of the underlying SAT problem. Other applications for which fair sampling is important include ground state entropy calculations, counting problems \cite{gomes2006model, gopalan2011fptas}, and machine learning \cite{hinton2002training, eslami2014shape}.

``Vanilla'' quantum annealing involves evolution with respect to a Hamiltonian
\begin{equation}
    H(t) = (1-t/T)\hd + (t/T)\hp, \qquad t \in [0, T],
\end{equation}
where $T$ is the total evolution time, $\hp$ is the `problem' Hamiltonian, diagonal in the $z$-basis, whose ground states are sought, and $\hd$ is a `driver' Hamiltonian capable of inducing transitions between $z$-eigenstates (usually $-\sum \sigma^x_i$). For sufficiently large $T$, the adiabatic theorem \cite{ambainis2004elementary} promises that the instantaneous ground state of $H(t)$ is approximately tracked throughout the evolution, so that a $z$-basis measurement at $t=T$ should return a ground state of $\hp$.

Perturbation theory arguments, simulations of small toy models, and actual QA hardware experiments on relatively large 2D square lattice spin models indicate that this simple protocol fails to sample degenerate ground states fairly \cite{matsuda2009ground,albash2015consistency, boixo2013experimental, konz2019uncertain, king2016degeneracy, zhang2017advantages, mandra2017exponentially}. In such cases, some of the ground states are ``hard'' suppressed (the probability of observing these states is approximately zero). Other cases feature ``soft'' suppression: all ground states are observed with non-zero probability, but some are seen more frequently than others. Soft suppression is usually repairable with a sufficient number of annealing runs and post-processing, but hard suppression is particularly detrimental for certain applications, such as those mentioned above. 

In principle, a denser transverse driver can mitigate the sampling bias, but such drivers are difficult both to engineer, and to simulate classically \cite{matsuda2009ground, konz2019uncertain}. As observed in \cite{konz2019uncertain}, even dense drivers cannot completely remove sampling bias, except in the extreme case of a complete graph driver. Another recent proposal \cite{yamamoto2020fair} based on ``extended'' quantum annealing \cite{somma2007quantum} also appears to be difficult to realize experimentally. 

In this note, we use perturbation theory arguments similar to those used in \cite{konz2019uncertain} to suggest a simple solution to the fair sampling problem, based on reverse annealing \cite{revann}. In particular, we consider random diagonal perturbations of the final Hamiltonian, which in perturbation theory trivially break degeneracy, leading to a uniform choice of computational basis state from the originally degenerate subspace in each run. We verify that these perturbation theory arguments are borne out via simulations of the small systems described in \cite{konz2019uncertain}. Though we do not take into account time-dependent effects in the perturbative arguments, we believe  that their inclusion should only serve to mitigate the problem further (via late-time transitions to excited states).

\emph{Perturbation theory}: Let $s:=t/T$ be a dimensionless time parameter. Towards the end of the anneal, where $1-s \ll 1$ (or rather when $s \approx 1-\lambda$ with $\lambda \ll 1$), $H(1-\lambda)$ can be viewed as a perturbation of the problem Hamiltonian $\hp$ by the driver $\hd$\footnote{At this stage we are assuming exact adiabatic evolution and ignoring time-dependent effects.}.
\begin{equation}
    H(1-\lambda) = \hp + \lambda \hd + O(\lambda^2)
\end{equation}
If $\hp$ has $m$ degenerate (computational basis) ground states $\kett{g_i}, i \in \{1,\dots,m\}$ with energy $E^{(0)}$, switching on the perturbation will result in the splitting of those states into $m$ eigenstates of $H(1-\lambda)$ with distinct energies\footnote{If the degeneracy is not completely lifted by the perturbation at first order, some of these eigenstates will share the same energy, but in the protocol described below this is rare.}. Perturbation theory posits that the energies $E_k(\lambda)$ and eigenstates $\kett{k}_{\lambda}, k \in \{1,\dots, m\}$ of $H(1-\lambda)$ can be expressed as a power series in $\lambda$:
\begin{align}
    \label{eq:2}
    E_k(\lambda) &= E^{(0)} + \lambda E^{(1)}_k + O(\lambda^2) \\
    \kett{k}_{\lambda} &= \kett{k^{(0)}} + \lambda \kett{k^{(1)}} + O(\lambda^2)
\end{align}

Inserting these Ans\"atze into the (time-independent) Schr\"odinger equation, one can compute corrections to the states and energies order by order in $\lambda$ in terms of quantities known from the unperturbed system \cite{Zwiebach}. For the resulting analysis to remain consistent, one must identify a ``good'' basis for the degenerate ground state subspace that diagonalizes a given perturbation. More specifically, suitable $\beta^i_k$ that define good basis states $\kett{k^{(0)}} = \sum_{i=1}^{m}\beta^i_k \kett{g_i}$ must be determined. $\beta^i_k$ can be obtained by solving the eigenvalue problem $W \beta_k = \epsilon_k \beta_k$ with $W_{ij} := \braa{g_i} V \kett{g_j}$. Here, eigenvalues $\epsilon_k$ provide the first order energy corrections $E^{(1)}_k$, the smallest of which enters the instantaneous ground state energy $E_{\tilde{k}} = E^{(0)} + \lambda E^{(1)}_{\tilde{k}}$ of $H(1 - \lambda)$ (where we have designated $\tilde{k} := \mathrm{argmin}_k \epsilon_k$).    

As pointed out in \cite{konz2019uncertain}, perturbation theory provides a simple explanation as to why some ground states are hard-suppressed in QA: Assuming exact adiabaticity, the corresponding eigenstate $\kett{\tilde{k}} = \sum \beta^i_{\tilde{k}} \kett{g_i}$ dictates the sampling probabilities in the computational basis; in particular, $\beta^i_{\tilde{k}} = 0$ leads to hard suppression of the $i^\mathrm{th}$ degenerate ground state of $\hp$ since the state being tracked by adiabatic evolution has no support on $\kett{g_i}$ near the end of the anneal. This suppression is caused by the sparse nature of the standard hypercube driver $V = -\sum{\sigma^x}$ as perturbation, and the set of ground states of $\hp$. Occasionally one can get lucky when using $V$, with all ground states reachable from each other by single bit flips, in which case soft suppression is a worst-case scenario. But this is not the case in general, even with denser drivers---only for the complete graph driver are we guaranteed that the late-time instantaneous ground state has support on all ground states of $\hp$. Unfortunately, engineering anything close to the complete graph driver seems prohibitively difficult. This compels us to search for alternative drivers that provide support for a would-be suppressed $\kett{g_i}$, at least in a perturbative analysis, if we seek to cure this suppression. 

\begin{figure}[H]
    \centering
    \includegraphics[scale=0.22]{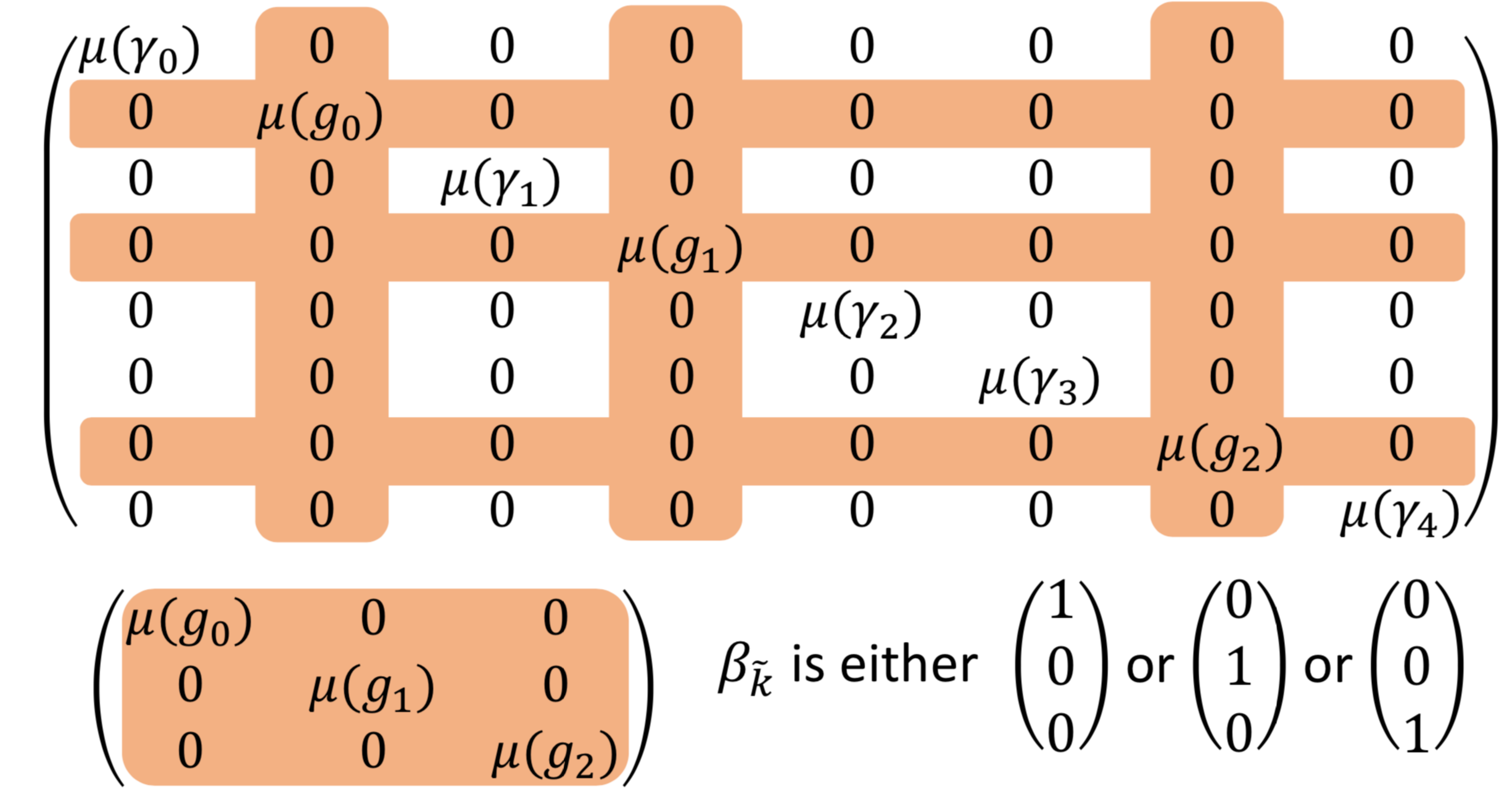}
\caption{\label{fig:1}Schematic showing $\hz$ for the $m=3$ case. Highlighted rows indicate the subspace spanned by the ground states $\{\kett{g_1}, \kett{g_2}, \kett{g_3}\}$. The reduced matrix (bottom left) in the ground state subspace has eigenvalues which supports either of the ground states.}
\end{figure}

\emph{Diagonal perturbations}: We argue here that adding diagonal Hamiltonians of the `reverse annealing' type to the driver can mitigate biased sampling. In particular, we consider
\begin{equation}
    \hz = \sum_i c_i \sigma_i^z
\end{equation}
with $c_i$ drawn randomly from a normal distribution. Since $\hz$ is diagonal, its eigenstates are computational basis states, and if $\hz$ were the only term to $O(\lambda)$ in the perturbed Hamiltonian, then the good basis that diagonalizes $\hz$ in the ground state subspace of $\hp$ is simply composed of the degenerate computational basis ground states $\kett{g_i}$ themselves. It follows that the $\kett{g_i}$ with the smallest $\hz$ eigenvalue will be favored in measurements. As we describe further below, running multiple anneals, each with a different perturbation characterized by randomly chosen $c_i$, will encourage each ground state of $\hp$ to be measured with roughly equal probability.

\emph{Properties of $\hz$}: For $c_i$ picked randomly from a normal distribution with zero mean and unit variance, in the ground state subspace spanned by some of the $\kett{g_j}$ (see Fig \ref{fig:1}), the corresponding eigenvalues are
\begin{equation}
    \mu(g_j) = \sum_{i=1}^n (-1)^{g_j(i)} c_i    
\end{equation}
where $g_j(i)$ represents the bit value of $\kett{g_j}$ at the $i^{\mathrm{th}}$ position. Since each eigenvalue is a linear combination of the $c_i$, they are also normally distributed, and across many instantiations, $\mathrm{argmin}_j \mu(g_j)$ is distributed uniformly; that is, a preferred degenerate computational basis state will be chosen uniformly at random in each annealing run to dominate the support of instantaneous ground state near the end of the anneal. In order to observe all $m$ degenerate ground states, an expected $\Theta(m \log m)$ runs are required.\footnote{See the ``coupon collector's problem.'' In the appendix we discuss the situation in which the number of ground states is not known in advance.}

\emph{Reverse annealing}: Feasibly engineerable time-dependent Hamiltonians that feature terms like $\hz$ as perturbations of $\hp$ at late times are present in so-called reverse annealing schemes \cite{perdomo2011study,ohkuwa2018reverse,yamashiro2019dynamics}, which fall under a broader class of protocols that employ non-convex combinations of initial and final Hamiltonians. Here we consider Hamiltonians of the type 
\begin{equation}
    \label{ranneal}
    H(s) = D(s)\hd + s\hp + Z(s)\hz,
\end{equation}
where again $\hd = -\sum_i{\sigma_i^x}$ is the standard hypercube driver, $\hp$ is the problem Hamiltonian, and $D(s)$ and $Z(s)$ are schedules which we specify further below. For our perturbation theory arguments, it is only necessary that
\begin{align}
    D(1-\lambda) & \approx O(\lambda^3), \label{driver_end} \\
    Z(1-\lambda) & \approx O(\lambda).
\end{align}
The reverse annealing paradigm would further specify
\begin{align}
    D(\lambda) & \approx O(\lambda), \\
    Z(\lambda) & \approx 1 - O(\lambda),
\end{align}
initializing the system in the ground state of $\hz$. The ground state $\kett{\gamma_{\mathrm{min}}}$ of $\hz$ has eigenvalue
$\mu(\gamma_{\mathrm{min}}) = \sum_{i=0}^{n-1} (-1)^{\gamma_{\mathrm{min}}(i)}c_i$ where $(-1)^{\gamma_{\mathrm{min}}(i)}c_i/\left|c_i\right| = -1$ for all $i$. These late-time conditions guarantee that the first-order perturbative corrections to the $\hp$ eigenstates vanish\footnote{\label{driver_note}This can also be relaxed to $D(1-\lambda)\approx O(\lambda^2)$ resulting in $O(\lambda)$ rotations of instantaneous eigenstates in the $\hp$ ground state subspace, which does not affect our general arguments.}, but (except in very rare cases) the eigenvalues are split, resulting in a single $\hp$ ground state providing the entire support of the instantaneous eigenstate.

Though perturbation theory is rather limited in what it can reliably say about the entire evolution (or the general utility of quantum annealing), its conclusions are corroborated by the simulated models considered here, which we describe below.

\begin{figure}[H]
    \includegraphics[scale=0.40]{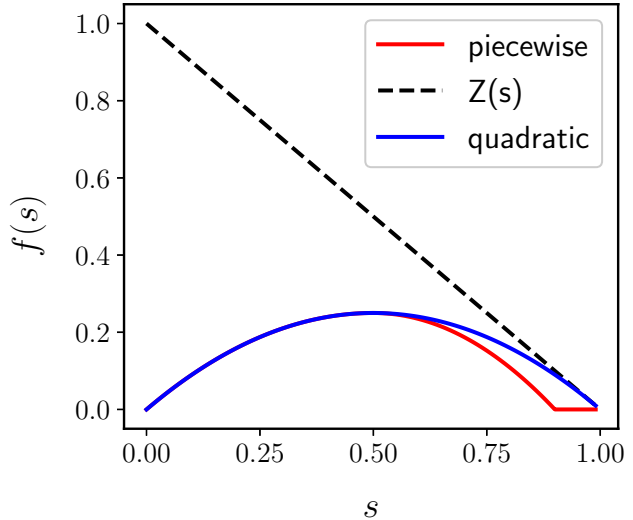}
    \caption{\label{fig:2}Schedules considered. The red and blue lines indicate the transverse field piecewise and quadratic driver schedules, respectively. $Z(s)$ is shown in black.}
    \label{fig:schedule}
\end{figure}

For simulations, we fix the schedule $Z(s) = 1 - s$, and consider two schedules for the transverse field. These are shown in Figure \ref{fig:2}; $\Ds(s)=s(1-s)$ is the quadratic driver schedule considered in \cite{perdomo2011study}, which violates (\ref{driver_end}) (and the looser condition in Footnote \ref{driver_note}), but allows us to examine situations where a linear combination of $\hd$ and $\hz$ plays the role of the perturbation. Because the good basis is generally not a subset of the computational basis in this case, the perturbation theory arguments are not as trivial, but if the coefficient of $\hz$ is sufficiently larger than that of $\hd$, then our conclusions should still hold approximately.

Contrast this with the piecewise schedule
\begin{equation}
  \Dpw(s) =
  \begin{cases}
  s(1 - s), &  0 \le s < 0.5 \\
  -\frac{25}{16}s^2 + \frac{25}{16}s - \frac{9}{64}, \quad  &  0.5 \le s < 0.9 \\
  0, & 0.9 \le s \le 1.0
  \end{cases}
  ,
\end{equation}

which ensures rather ham-handedly that the transverse field is absent in the perturbation. In this case we expect sampling probabilities to be solely dictated by $\hz$. 

\begin{figure}[H]
    \includegraphics[scale=0.30]{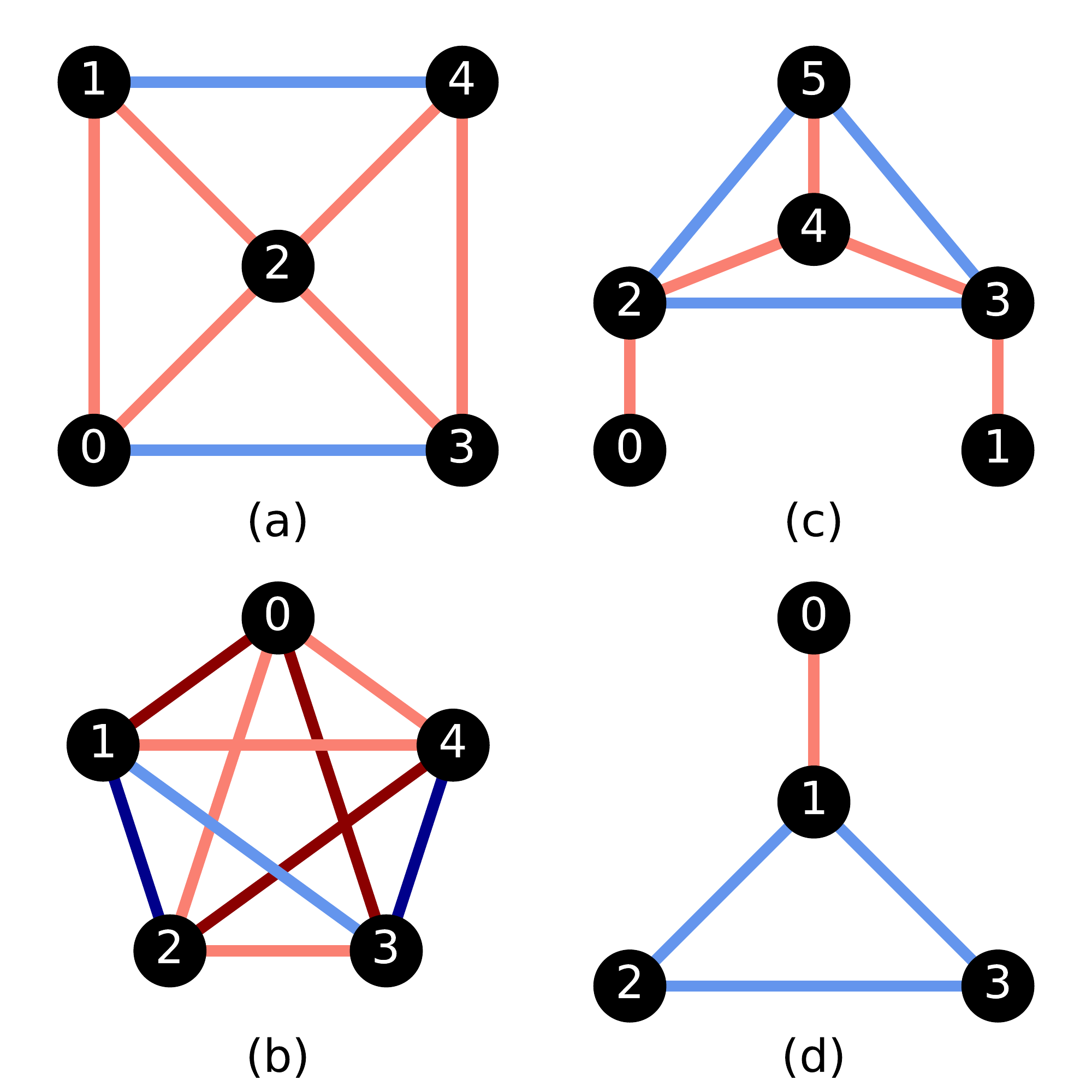}
    \caption{\label{fig:3}The instances we consider. The coupling values are either 2 (dark red edges), 1 (orange edges), $-1$ (light blue edges), or $-2$ (dark blue edges).}
    \label{fig:models}
\end{figure}

\emph{Small instances}: Here we revisit the particular instances studied in \cite{konz2019uncertain} and shown in Figure \ref{fig:models}. The problem Hamiltonians are zero-bias Ising models
\begin{equation}
    \hp = -\sum_{(i,j)}J_{ij}\sigma^z_i\sigma^z_j
\end{equation}
and are depicted graphically in Figure \ref{fig:3}, where each node corresponds to a qubit and each edge to a non-zero coupling $J_{ij}$. Consulting the figure, the coupling values are either 2 (dark red edges), 1 (orange edges), $-1$ (light blue edges), or $-2$ (dark blue edges). (a)-(c) have been shown to exhibit hard suppression while (d) features soft suppression.

\emph{Simulations}: The perturbative arguments we have made so far only hold in the adiabatic limit. When time-dependence is taken into account and spectral gaps necessarily close, diabatic transitions will be induced. Though we do not provide an analytic argument, these transitions would presumably only further democratize sampling from the ground state subspace given the convergence of eigenvalues at late times. Indeed, the numerical simulations performed here appear to substantiate this intuition.

\begin{figure}[H]
    \includegraphics[scale=0.38]{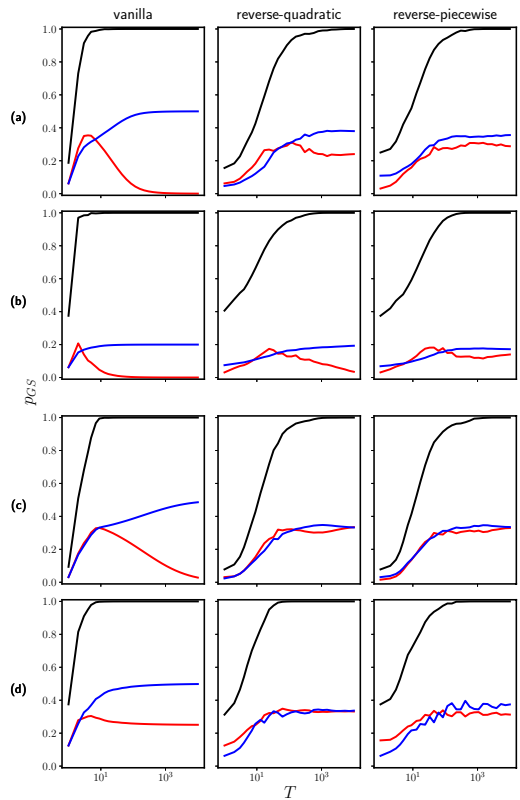}
    \caption{\label{fig:4}Measurement probability versus total annealing time $T$ for each of the ground states and schedules. Reverse annealing measurement probabilities were averaged over 32, 32, 64, and 16 trials for the models (a), (b), (c), and (d), respectively, where each trial corresponds to a randomly generated $\hz$. The labels correspond to the model labels in Figure \ref{fig:3}. Red and blue curves show (up to symmetry) the probability of individual $\hp$ ground states. The black curves show the sum of probabilities over all ground states.}
    \label{fig:probabilities}
\end{figure}

In Figure \ref{fig:probabilities}, we plot squared magnitude of each $\hp$ ground state coefficient as a function of the total annealing time $T$, obtained by integrating the Schr\"odinger equation in Qutip \cite{johansson2013qutip, johansson2012qutip}. For the reverse annealing runs, we plot the average of these probabilities over multiple trials, where each trial corresponds to a randomly generated $\hz$, and we observe that this protocol does improve the sampling bias. In the case of the piecewise-defined schedule $\Dpw$ where $\hz$ is the sole first-order perturbation to $\hp$ at late times, we see that for all models, states that were suppressed by vanilla annealing now provide roughly equal support to the wave function for large enough $T$. We also observe that the quadratic driver schedule $\Ds$, in which both $\hd$ and $\hz$ contribute at first order, is less effective at removing the bias in models (a) and (b).

\begin{figure}[H]
    \includegraphics[scale=0.6]{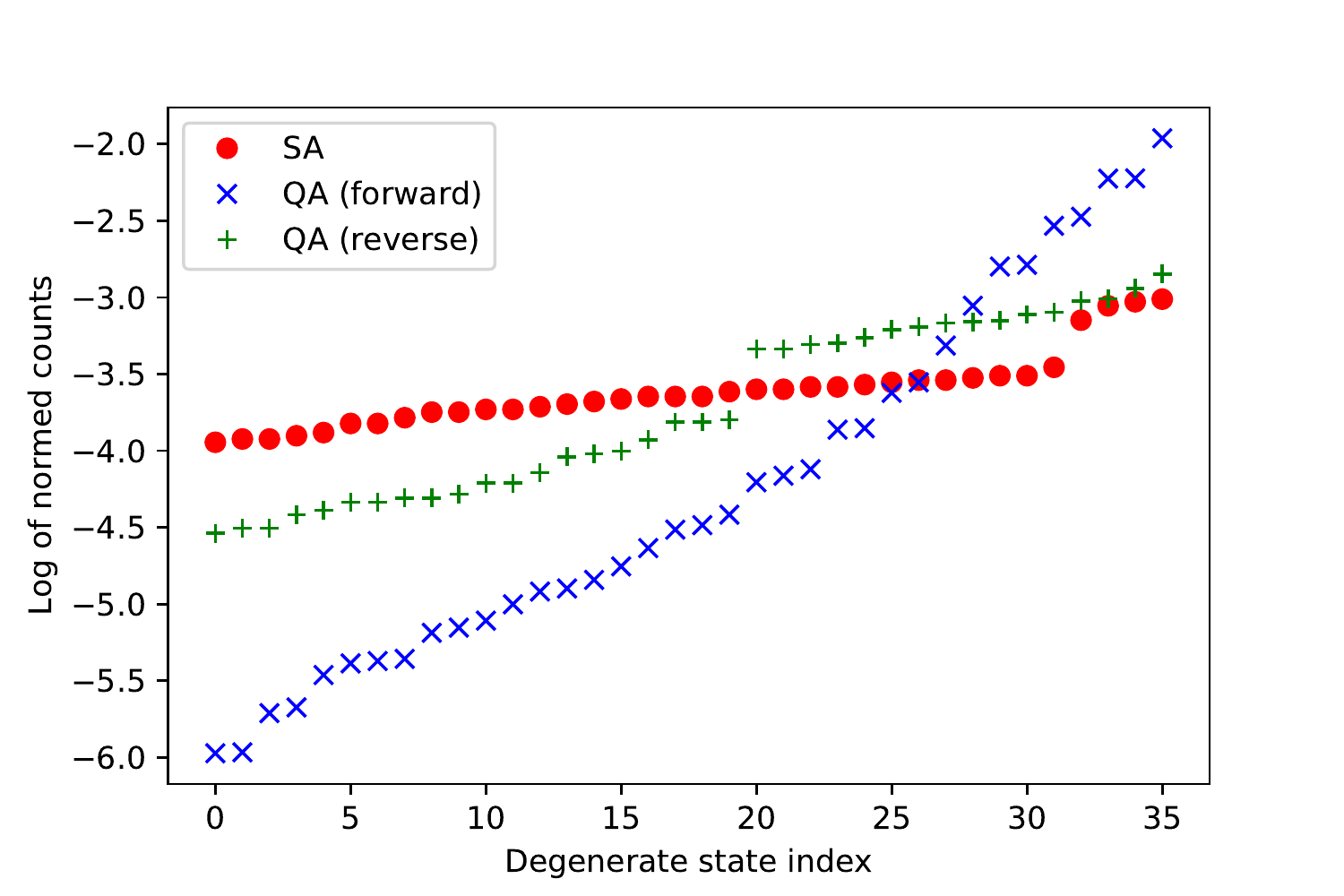}
    \caption{\label{peg}Log of normed counts for each solution method. The horizontal axis simply enumerates the 36 ground states associated with this instance, and vertical axis corresponds to the (log of the) number of times that ground state was observed relative to the other ground states.}
    \label{peg_fig}
\end{figure}

\emph{Hardware runs}: We have also examined larger spin glass instances supported on graphs native to D-Wave's Chimera (using 160 qubits) and Pegasus (using 92 qubits) architectures. As in \cite{mandra2017exponentially}, we consider instances with couplings randomly drawn from $\{\pm1, \pm2, \pm4\}$. For an initial pool of these instances, individual couplings are adjusted to keep the number of degenerate ground states manageable by eliminating so-called free spins (sites whose total local field vanishes; each of these doubles degeneracy)\footnote{This fairly expensive process was the main constraint on the size othe instances we considered.}. Then simulated annealing (SA) and parallel tempering with iso-energetic cluster moves (PT-ICM) (see Appendix \ref{classical_samp}) were used to find the ground states of each instance, and those with the largest number of degenerate ground states were selected for additional classical analysis and quantum annealing runs. The latter were carried out on LANL's D-Wave 2000Q system, and D-Wave's Advantage system in Burnaby. To the best of our knowledge, these machines do not allow a user to explicitly program the initial Hamiltonian (here $H_z$), but the same effect is achieved by programming an initial state (the would-be ground state of $H_z$), which is supported via reverse annealing functionality, and we simply specify uniformly chosen random initial states in these experiments. Outputs of the classical algorithms and quantum annealer runs were averaged over eight blocks of samples of size $500$, obtained from independent trials, and we observed similar behavior in all of the instances and on both annealing devices. Results from an example instance based on a native Pegasus subgraph with 36-fold degeneracy are shown in Figure \ref{peg_fig}.

As expected, vanilla QA exhibits a substantial sampling bias (note that the vertical axis is log-scaled), while our reverse annealing protocol performs comparably to simulated annealing. We point out however that, not surprisingly, the magnitude of the driver term (here controlled by an overall multiplicative factor of $\Dpw$) plays an important role in the success or failure of our protocol; too large, and one is essentially back to vanilla annealing and the associate bias shows up; too small, and only a very small region near the initial state is explored (and indeed many ground states are never observed). In these experiments, we found that setting the driver term to $2 \Dpw \hd$ gave good results.

In the more realistic scenario where one does not know the dimension of the ground state subspace \emph{a priori}, we give an algorithm in Appendix \ref{algo} that will return the full set of ground states (up to some specified failure tolerance), assuming the annealing protocol described here uniformly samples only the ground states. This latter assumption is of course dubious for contemporary devices, so additional multiplicative overhead is usually required.

\emph{Remarks}: We have presented a very simple protocol, motivated by time-independent perturbation theory, that is able to cure a seemingly hard-wired tendency towards biased sampling afflicting vanilla quantum annealing. While it is well-known that the complete graph driver ensures fair sampling, no sparser driver can provide this guarantee in general, and in any case denser drivers seem extremely difficult to realize experimentally. On the other hand, since our method is very similar to existing reverse annealing techniques that have experimental realizations, it is also a practical solution to the unfair sampling problem. 

\section{Acknowledgements}
We thank Michael Jarret for scathing comments on an initial draft and fruitful discussions, and D-Wave and Joel Gottlieb in particular for providing access to the Advantage system via their Beta program.

\bibliography{BAB}

\appendix
\section{Classical sampling schemes}
\label{classical_samp}
Simulated annealing runs were carried out using D-Wave's ``neal'' library functionality, and the PT-ICM Monte Carlo runs were performed using custom code. Two spin replicas at each $\beta$ and $20$ replicas were maintained at each $\beta$. A total of $N_{\mathrm{sweep}} = 10^{14}$ MC sweeps were made, where each sweep consisted of: a) $N$ single flips for a randomly selected spin replica at each $\beta$; b) an iso-energetic clustering move between spin replicas at a randomly chosen $\beta$; and c) parallel tempering swaps between a randomly chosen pair of $\beta$ replicas. After $N_{\mathrm{sweep}}/2$ sweeps, if both the spin replicas at the highest $\beta$ were seen at the same energy, then that energy was recorded as the lowest energy. If a lower energy was observed after $> N_{\mathrm{sweep}}/2$ sweeps, then the lowest energy was updated. Final samples at the highest $\beta$ were obtained at the end of the run. For both SA and PT-ICM, a geometric $\beta$ schedule was used, where the $\beta$ range was set to $[\frac{1}{3.05}, \frac{1}{0.05}]$ with a bin size of $1.22$.

\section{An algorithm}
\label{algo}
We give an algorithm for generating (up to some failure rate) the set of all degenerate ground states $G$, assuming each is returned uniformly at random by the annealing procedure described in the main text. The key quantity is an estimate of the number of trials required to guarantee with high probability that all ground states have been observed. This is equivalent to the problem of determining the total number $m$ of coupons in the coupon collector's problem. For fixed $m$ we have the following bound for the number of trials $T$ required to collect all $m$ coupons:
\begin{equation} \label{cc}
    P[T>m\log (m/\epsilon)] \leq \epsilon
\end{equation}
The algorithm below is composed of possibly several rounds of sampling, each round corresponding to an assumed $m$, and we use this inequality to bound the failure rate of the entire procedure. Specifically, if one asks for a success rate of $1-\epsilon$ after $r$ rounds of sampling, one needs
\begin{equation}
    T \ge \lceil m\log (rm/\epsilon) \rceil =: T(m,r,\epsilon)
\end{equation}
samples per round. By iteratively doubling a guess for $m$ as unseen states are sampled, we have that the maximum number of rounds is $n-1$ (for an $n$-qubit system; we begin with $m=2$). 
\renewcommand{\algorithmicrequire}{\textbf{Input:}}
\renewcommand{\algorithmicensure}{\textbf{Output:}}

\begin{algorithm}[H]
    \caption{GetGroundStates}
    \begin{algorithmic}[1]
        \Require System size $n$, failure tolerance $\epsilon$ 
        \Ensure Set of ground states $G$
        \State $m \gets 2$
        \State $T \gets T(m,n,\epsilon)$
        \State $G \gets \emptyset$
        \State $t \gets 0$
        \While{$t \leq T$}
            \State $g \gets$ \textproc{AnnealOnce()}
            \State $G \gets G \cup \{g\}$
            \State $t \gets t + 1$
            \If{$|G| > m$}
                \State $m \gets 2m$
                \State $T \gets T(m,n,\epsilon)$
                \State $t \gets 0$
            \EndIf
        \EndWhile
        \State \Return $G$
    \end{algorithmic}
\end{algorithm}

\end{document}